# Expected performance of a hard X-ray polarimeter (POLAR) by Monte Carlo Simulation

Shaolin Xiong[a,1], Nicolas Produit[b], Bobing Wu[a]

[a]Key Laboratory of Particle Astrophysics, Institute of High Energy Physics, Chinese Academy of Sciences, Beijing, China

[b]ISDC, Université de Genève, Switzerland

**Abstract**

Polarization measurements of the prompt emission in Gamma-ray Bursts (GRBs) can provide diagnostic information for understanding the nature of the central engine. POLAR is a compact polarimeter dedicated to the polarization measurement of GRBs between 50-300 keV and is scheduled to be launched aboard the Chinese Space Laboratory about year 2012. A preliminary Monte Carlo simulation has been accomplished to attain the expected performance of POLAR, while a prototype of POLAR is being constructed at the Institute of High Energy Physics, Chinese Academy of Sciences. The modulation factor, efficiency and effective area, background rates and Minimum Detectable Polarization (MDP) were calculated for different detector configurations and trigger strategies. With the optimized detector configuration and trigger strategy and the constraint of total weight less than 30 kg, the primary science goal to determine whether most GRBs are strongly polarized can be achieved, and about 9 GRBs/yr can be detected with MDP < 10% for the conservative detector configuration.

## 1 Introduction

Polarization measurements in X-ray and gamma-ray domain by means of the photoelectric effect, Compton scattering or pair production is one of the most exciting frontiers in contemporary astrophysics, since it intends to serve as a potential diagnostic approach to discriminate models which could be very different but can explain successfully observations except for polarization. Photons emitted from celestial objects contain four kinds of information: energy, direction, time, and polarization. Although energy, direction and time are measured as spectrum, image and intensity, polarization has rarely been exploited in the X-ray and gamma-ray band.

Almost all non-thermal emission mechanisms creating X-ray and gamma-rays can produce high degrees of linear polarization which also depend on the magnetic field and the geometry of the source emission zone [1]. All magneto-bremsstrahlung radiation, including cyclotron, synchrotron and curvature radiation, are potential sources of linearly polarized photons whose polarization degree depends on the configuration of the magnetic field. For the observed range of power-law indices, from 1.5 to 5.0 for astrophysical synchrotron radiation sources, the maximum observed degree of linear polarization is expected to vary from approximately 65% to 80%, which will be reduced by the inhomogeneity of the magnetic field configuration [1]. Electron-proton bremsstrahlung radiation can produce linear polarization as high as 80%.

---

Compton scattering can create energy-independent polarization [1, Equation (2.31), (2.33)], which is opposite to the synchrotron radiation and thus could be used to distinguish one from the other. Magnetic photon splitting can also lead to polarization levels of up to 30% [1]. Furthermore, before the polarized photons reach the Earth, the environment during their journey will also mark its signature in the polarization information, for instance, Compton scattering or Faraday rotation can change the status of polarization. Therefore, polarimetry in high energy regime is expected to yield crucial information about emission mechanism, geometry, magnetic field and environment in the journey for a wide variety of high energy astrophysical sources, such as GRBs, soft gamma repeaters (SGRs), solar flares, isolated pulsars, jet-dominated AGNs, accreting black holes and neutron stars [2,3].

Although polarimetry is such an important and powerful tool, it has rarely been implemented successfully and accurately in X-ray or gamma-ray range, especially in the energy band above 30 keV. A sounding rocket launched in year 1971 was first to measure X-ray linear polarization in the Crab Nebula, which determined that the X-ray emission in the nebula was due to synchrotron radiation [4]. OSO-8 searched for linear polarization for 15 bright X-ray sources with a Bragg crystal polarimeter. For most sources the polarization were of low significance [5], however, observations of the Crab Nebula showed a polarization of 19.2 ± 1.0% at 2.6 keV and 19.5 ± 2.8% at 5.2 keV when contamination from the pulsar was removed [6], for Cygnus X-1 a polarization of 2.4 ± 1.1% at 2.6 keV and 5.3 ± 2.5% at 5.2 keV [7] and for Scorpius X-1 a polarization of 0.39 ± 0.20% at 2.6 keV and 1.31 ± 0.40% at 5.2 keV [8]. COMPTEL [9], BATSE [10], both aboard the CGRO, attempted to gain some polarization signature from their data. However, no successful results have so far been obtained from COMPTEL [9]. Although some analysis of the BATSE Albedo Polarimetry System (BAPS) data shows strong evidence that the lower limits of polarization degree are 35% and 50% in the prompt flux of GRB 930131 and GRB 960924, respectively, the polarization degree cannot be firmly constrained beyond a systematic error [11]. RHESSI [12], a solar X-ray and gamma-ray spectroscopic imaging detector as well as a possible hard X-ray polarimeter [13] has measured 80 ± 20% linear polarization of the prompt emission in the energy range 15-2000 keV of the GRB 021206 [14], which has not been confirmed by the other two independent studies of the same data [15,16], so that the degree of polarization for GRB 021206 remains uncertain. In principle, both the IBIS and SPI instruments on INTEGRAL were capable of polarimetry in the gamma-ray range [17,18]. But the capabilities of IBIS were limited by the high level of random coincidences between the two detection planes and by the geometry, which was not optimal. Nevertheless, SPI has measured a high level of polarization of the very intense burst GRB 041219a, but the systematic effects which could mimic the weak polarization signal couldn't be excluded, hence this result cannot significantly constrain GRB models [19,20]. However, SPI has been reported to detect gamma rays of 46±10% polarization from the vicinity of the Crab pulsar recently [21]. The detector plane of the BAT onboard Swift [22,23] may have worked as a good polarimeter, but the design of both detectors and signal processing electronics are not suitable for selecting the Compton scattering events [2].

In order to actualize polarimetry to various cosmic sources in the X-ray and gamma-ray energy range, many polarimeters have been proposed to this date: Photoelectric polarimeters based on a micropattern gas chamber (MPGC) for X-ray Evolving Universe Spectroscopy (XEUS) and POLARIX [24,25,26], Compton polarimeters like the Gamma-Ray Polarimeter (GRAPE) [27], the Polarized Gamma-ray Observer (PoGO) [28] and the smaller version PoGOLite [29], the soft gamma-ray detector (SGD) onboard NeXT [30], the Coded Imager and Polarimeter for High Energy Radiation (CIPHER) [31], the X-ray Polarimeter Experiment (XPE) [32], as well as POLAR [33], and Pair production polarimeters [34].

## 2 Compton Scattering Polarimetry

For a polarized photon having undergone the Compton scattering (see Fig. 1), the differential cross section is given by the Klein-Nishina formula:

$$\frac{d\sigma}{d\Omega} = \frac{r_0^2 \varepsilon^2}{2}\left(\frac{1}{\varepsilon} + \varepsilon - 2\sin^2\theta\cos^2\eta\right)$$

$$= \frac{r_0^2 \varepsilon^2}{2}\left(\frac{1}{\varepsilon} + \varepsilon - \sin^2\theta + \sin^2\theta\cos\left(2\left(\eta + \frac{\pi}{2}\right)\right)\right)$$

(1)

where $r_0$ is the classic radius of electron and $\varepsilon = E'/E$. For a fixed scattering angle, the cross section reaches the maximum at $\eta = \pi/2$ and the minimum at $\eta = 0$, and the azimuthal distribution of the scattered photons follows a $\cos(2\eta)$ distribution (see Fig. 2).

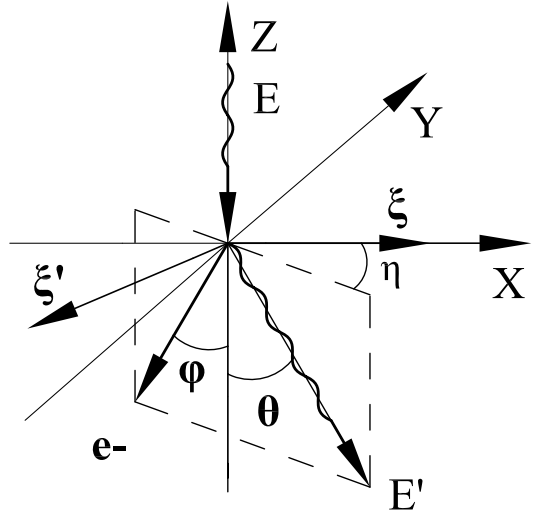

*Fig. 1. Compton scattering of the polarized photons. E and $\xi$ are the energy and electric vector of the incident photon. $E'$, $\xi'$ and $\theta$ are the energy, electric vector and scattering angle of the scattered photon. $\varphi$ is the recoil angle of the electron. $\eta$ is the azimuthal angle which represents the angle of the plane of scattered photon and recoiled electron with respect to $\xi$.*

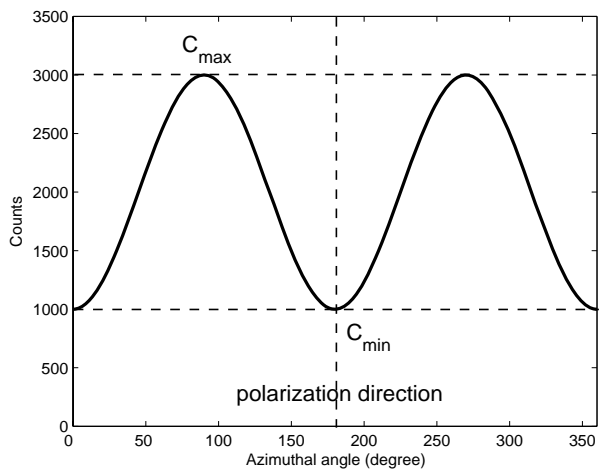

*Fig. 2. The azimuthal distribution of scattered photons. The central vertical dashed line represents the polarization direction.*

In order to use a polarimeter, the response of the polarimeter to the 100% polarized photons must be calibrated. This response is known as the modulation factor, M, which is defined as:

$$M = \frac{C_{max} - C_{min}}{C_{max} + C_{min}} \quad (2)$$

where $C_{max}$, $C_{min}$ refer to the maximum and minimum numbers of counts in the azimuthal distribution. Since the azimuthal distribution will be uniform for an unpolarized photon beam, the polarization level of the beam is given by:

$$P = \frac{M_p}{M_{100}} \quad (3)$$

where $M_p$ is the modulation factor for the measured photon beam, $M_{100}$ is the modulation factor for the 100% polarized beam, and P is the polarization level of the beam. The Minimum Detectable Polarization (MDP), which is the minimum level of polarization that is detectable with a significance level, $n_\sigma$ (number of sigma), due to statistical variations, can be expressed as:

$$MDP = \frac{n_\sigma}{M_{100} S} \sqrt{\frac{2(S+B)}{T}} \quad (4)$$

where S is the total source counting rate, B is the total background counting rate and T is the observation time.

Generally, a Compton scattering polarimeter measures the azimuthal distribution of the scattering photons to analyze the polarization information, then two kinds of detectors are required to define the azimuthal angle, one is the scattering detector which is used as the target and determines the position of the Compton scattering, the other is the tracking detector which tracks the scattered photon by measuring the position of Compton scattering or photoelectric absorption between the scattered photon and tracking detector.

## 3 POLAR

Gamma-Ray Bursts, the mysterious brightest electromagnetic explosions in the universe, remain one of the most interesting topics in high energy astrophysics. Many competing theoretical models, like the fireball model, electromagnetic model and cannonball model, can explain the current observations in the energy and time domains, but these models make different predictions about polarization which hasn't been exploited explicitly in the X-ray and gamma-ray band. Thereby polarimetry will become one of the ultimate observables to distinguish different models. In this case, POLAR was proposed as a dedicated Compton polarimeter for GRBs [33,35].

The energy band of POLAR was selected to 50-300 keV, because it is the most sensitive band for the Compton polarimeter and sufficient photons of GRBs can be detected in this band. The basic design of POLAR is optimized for 50-300 keV and a wide field of view (FOV) in order to catch GRBs. The instrument is characterized by a large area and high analyzing power, which utilizes low Z, fast plastic scintillator bars read out by multi-anode photo-multiplier tubes (MAPMTs) arranged in a large array. Plastic scintillator bars are used as one-dimension position-sensitive detector for both scattering detection and tracking detection. This will greatly improve the detection efficiency at the cost of bad energy reconstruction compared to the design that utilizes light scintillator as scattering detector and heavy scintillator as tracking detector [27]. Monte Carlo modeling shows that systematics due to bad energy reconstruction does not play a big role in the case of a Band spectrum [36, and Equation (6)]; hence we think the gain in the efficiency is more important than the loss in energy resolution. In case of energy dependent polarization, POLAR will measure an average polarization over the energy range.

The schematic drawing of one POLAR detector unit is shown below (see Fig. 3). The plastic scintillator bars (BC-408) are 6 mm×6 mm square and 200 mm height, and 8×8 bars are coupled with one H8500 MAPMT. In order to enlarge the effective area, POLAR will consist of tens of detector units. Considering the very strong constraint of total mass less than 30 kg from the Chinese Space Laboratory, the ultimate design of POLAR cannot be fixed currently. Therefore two configurations of POLAR are simulated: 4×4 and 5×5 detector units. The front and lateral faces of the target are shielded by a 1 mm thick carbon fiber layer to absorb the very low energy photons and charged particles.

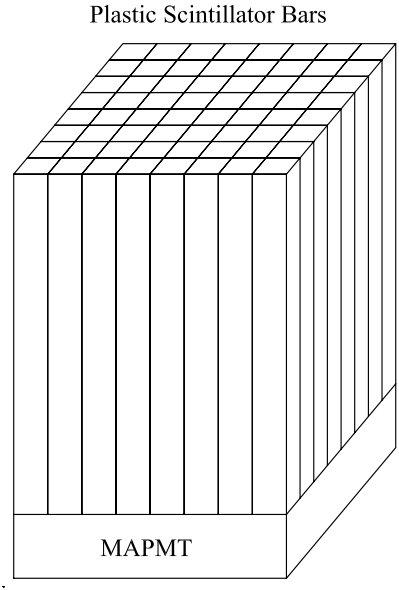

*Fig. 3. A detector unit of POLAR. The actual POLAR will be an array of detector units.*

## 4 Working principles

When a GRB occurs in POLAR's FOV, the count rate will increase rapidly and fulfill the criterions set in advance, producing a GRB trigger. Then all data recorded by POLAR (including some data before and after the GRB for background reduction) will be saved by the spacecraft and telemetered to the ground. As seen in Fig. 4, the gamma-rays will interact with plastic scintillator bars. According to the principle of Compton polarimetry, an effective event contains at least two coincident interactions. Although small angle scatterings dominate the Compton scattering, the direction of polarization of the incident photon is conserved basically and small angle scattering induce small energy deposition. The azimuthal angle can be reconstructed by the bars with the two highest energy depositions [33]. Because the light collection efficiency of the plastic bars employed is about 40% (based on simulations and experiments) and the quantum efficiency of the PMT is about 20%, an energy deposition of 5 keV will produce about 4 photo-electrons. Therefore, the minimum energy deposition threshold per bar is set to 5 keV.

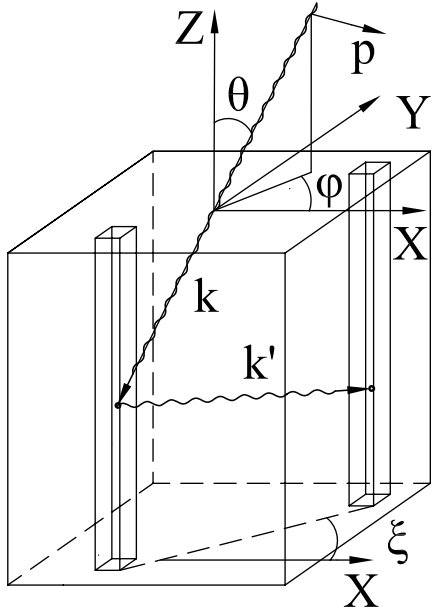

*Fig. 4. Geometry of the Compton scattering event. The azimuthal angle $\xi$ derived from the two bars where interactions occur. $\theta$, $\varphi$ are the entrance angle of the incident photons relative to the detector's fixed coordinate system (XYZ). $k$ is the direction of the incident photons and $p$ is the polarization direction.*

A charged particle traversing the detector will deposit a huge amount of energy in comparison to typical gamma-ray events, so a cut on total energy deposition is very efficient against cosmic rays. Cosmic rays also cause adjacent bars to fire because of continue energy loss. Therefore, to eliminate the background induced by those charged particles, the trigger should exclude the case that the two highest energy depositions occur in the adjacent bars. This cut will cause a reduction in the detection efficiency (< 10% [35]) and a very small portion (<< 1% based on simulations) of charged particles contributing to the background.

To summarize, the trigger strategy for effective events is that there are at least two coincident interactions with energy deposition larger than 5 keV, the two highest energy depositions occur in non-adjacent bars and the total energy deposition is less than 500 keV. The two non-adjacent bars with the biggest energy depositions define the azimuthal angle.

As it is difficult to localize the GRBs and measure their spectra by POLAR itself, the photon entrance angles and spectra should be known before estimating $M_{100}$. Fortunately, this information can be derived from other mission through the GCN (The Gamma ray bursts Coordinates Network [37]) and other astrophysics missions in orbit at the same time.

## 5 Simulations

As mentioned in Section 2, the MDP is the characteristic parameter for a polarimeter. Prior to calculating the MDP, the following parameters should be known:
a) Modulation factor,
b) Effective area,
c) Background,
d) Source intensity and observation duration.

The first three parameters can be derived by Monte-Carlo simulations utilizing the Geant4 toolkit [38], while the last one can be extracted from the BATSE Current Gamma-Ray Burst Catalog [39]. Moreover, considering the appearance probability of GRBs from this catalog, one can estimate how many GRBs with a certain MDP can be observed by POLAR in one year of observation.

The main sections of the simulation work are described briefly below:

First of all, construct a detector model of POLAR in Geant4 according to the design in Section 3. Two detector models are simulated, one consists of 4×4 detector units (denoted as DM1), for which most of results in the following text are based, while the other is composed of 5×5 detector units (denoted as DM2), which is probably a little over the mass budget but has not been confirmed yet.

Secondly, specify physical processes that are required in the Geant4 simulation. The most important one is the low energy Compton process for the polarized photon.

Then, set the trigger strategy for effective events. As discussed in Section 4, the trigger strategy here is that the two plastic scintillator bars defining the azimuthal angle are non-adjacent and have the two highest energy depositions out of all bars, these two highest energy depositions are not less than 5 keV, and the total energy deposition is less than 500 keV.

And then, set the properties of incident photons, including incident direction (in the following text, theta and phi refer to $\theta$ and $\varphi$ in Fig. 4), energy or spectrum, polarization level and polarization direction (denoted as polarization angle in the following text).

Finally, analyze the simulation data produced by Geant4, get the azimuthal distributions and parse the information for the modulation factor, effective area, and background rate.

A typical azimuthal distribution is showed in Fig. 5, which is derived from the simulation when the input spectrum is Band spectrum [36, and Equation (6)], the incident direction is theta=40 degree and phi=45 degree, and the detector model is DM1. The azimuth angle range from 0 degree to 360 degree is divided uniformly into 40 channels. Before the azimuthal distribution can be fitted properly using a cosine function (see Fig. 2), the asymmetry due to the geometry of the polarimeter should be removed using the formula below [1],

$$N_{true}(\phi) = \frac{N_{pol}(\phi)}{N_{non}(\phi)} N_{norm} \qquad (5)$$

where $N_{pol}$ is the azimuthal distribution of a certain polarization level (see Fig. 5), $N_{non}$ is the azimuthal distribution of non-polarization (see Fig. 6), $N_{norm}$ is a normalization factor, $N_{trus}$ is the true azimuthal distribution (see Fig. 7).

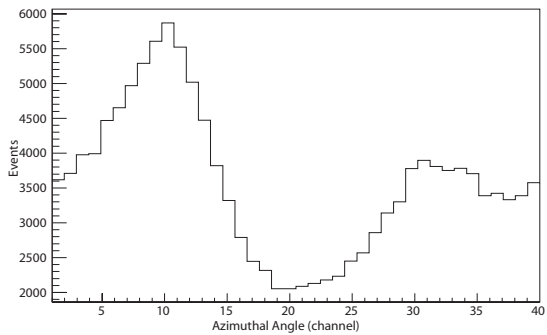

Fig. 5. The azimuthal distribution in the condition of theta=40 degree, phi=45 degree, polarization level=100%, DM1.

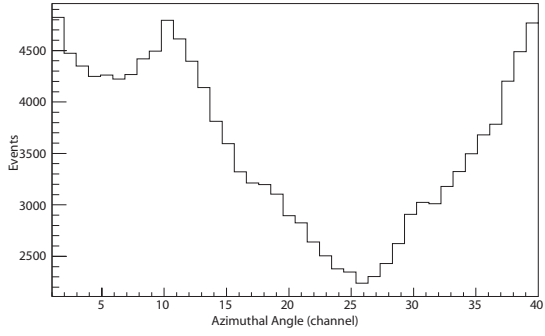

Fig.6. The azimuthal distribution in the condition of theta=40 degree, phi=45 degree, polarization level=0%, DM1.

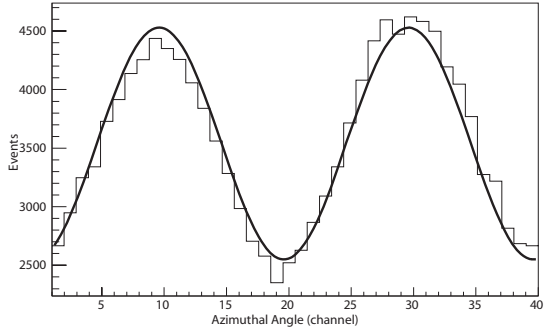

*Fig. 7. The true azimuthal distribution derived according to Equation 5 in the condition of theta=40 degree, phi=45 degree, polarization level=100%, DM1 can be fitted properly. The reduced $x^2 = 0.944$ for fit, and the modulation factor is 0.27.*

In fact, the asymmetry comes not only from the geometry of polarimeter but also from the non-uniformity of the detection of the whole polarimeter. In our case, the Space Laboratory won't allow POLAR to rotate, which is often used to compensate for the non-uniformity. Consequently, the asymmetry of POLAR will be calibrated appropriately on ground and in orbit. The high statistic from the unpolarized diffuse background averaged over one orbit will allow us to subtract the inhomogeneity effect. The performance of each bar will be tracked before and after the GRB events.

Fitting the asymmetry-reduced azimuthal distribution will give the modulation factor according to Equation 2, while integrating the counts of the azimuthal distribution will yield the efficiency compared to the input photons counts. The effective area is the product of the efficiency and geometric area of the detector.

5.1 Modulation factor and effective area

In this section, two very important properties of the polarimeter, modulation factor and effective area, will be investigated as a function of energy, incident direction (theta, phi) and polarization angle of incident polarized photons.

The relationship between modulation factor, effective area and polarized photon energy is shown in Fig. 8 and Fig. 9, respectively. Both the modulation factor and effective area reach a maximum when the energy is about 150 keV for the same incident angle. However, for the same photon energy and phi (phi = 90 degree), when theta varies from 45 degree to 0 degree, the modulation factor increases whereas the effective area decreases. For the same photon energy and theta (theta = 45 degree), when phi varies from 45 degree to 90 degree, the modulation factor increases whereas the effective area decreases. The reason is that for the oblique incidence case, the detection plane with position resolution is not perpendicular to the incident direction, then according to Fig. 1 the measured modulation is actually a partition of the real modulation, and the geometric area is the cross section area seeing from the incident direction, then the effective area is larger than that of on-axis (theta = 0 degree).

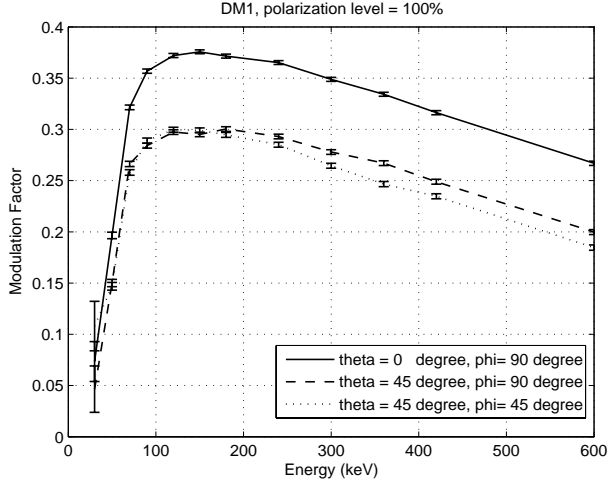

*Fig. 8. The modulation factor varies with the photon energy as well as the incident angle.*

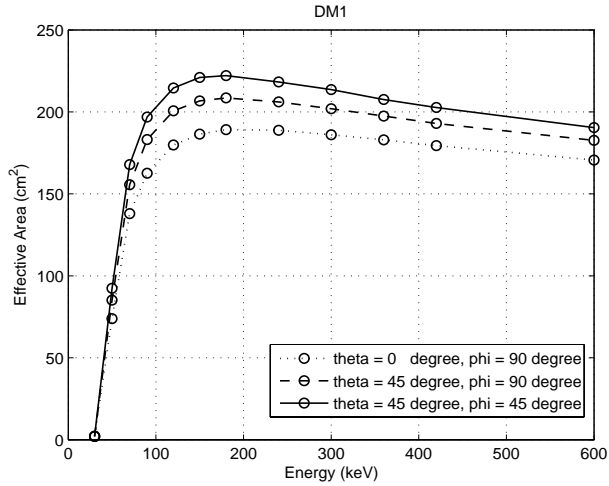

*Fig. 9. Effective area varies with the photon energy as well as the incidence angle*

Since the GRBs are the main objects that POLAR will detect, the energy spectrum of incident photons is set to the typical Band spectrum [36]:

$$f(E) = A(\frac{E}{100 keV})^\alpha \exp(-\frac{E}{E_0}), (\alpha-\beta)E_0 \geq E$$
$$= A\left[\frac{(\alpha-\beta)E_0}{100 keV}\right]^{\alpha-\beta} \exp(\beta-\alpha)(\frac{E}{100 keV})^\beta, (\alpha-\beta)E_0 \leq E \quad (6)$$
$$E_{peak} = (2+\alpha)E_0$$

where α = -1.01, β = -3.31, and $E_{peak}$ =390 keV.

All simulations hereafter are based on the Band spectrum with an energy range of 10 to 300 keV. The trigger strategy used hereafter eliminates the case that two adjacent bars with the two highest energy depositions define the azimuthal angle.

The data in Fig. 10 and Fig. 11 clearly shows the dependency of the modulation factor on the polarization angle, which hasn't been investigated in previous work [33,35]. The modulation factor changes periodically with the polarization angle because of the square geometry of POLAR, and this effect is especially notable when theta > 70 degree. Considering this effect and the unavoidable shielding structure around POLAR when installed on the Chinese Space Laboratory, the field of view (FOV) of POLAR will be set to 1.3π sr (theta varies from 0 to 70 degree and phi varies from 0 to 360 degree). Reducing the FOV will reduce the background and therefore enhance the sensitivity, but will reduce the number of detected GRBs. Consequently, the overall capability of polarimetry will not changed apparently with this FOV compared to the $2\pi$ sr FOV (theta varies from 0 to 90 degree).

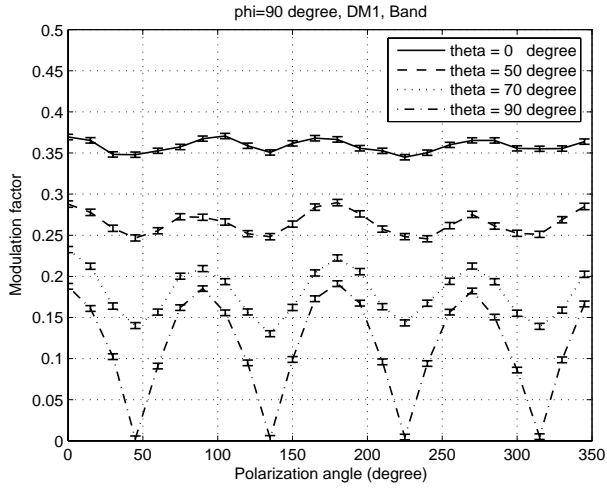

*Fig. 10 The modulation factor varies with the polarization angle of incident photons when the incidence angle phi=90 degree*

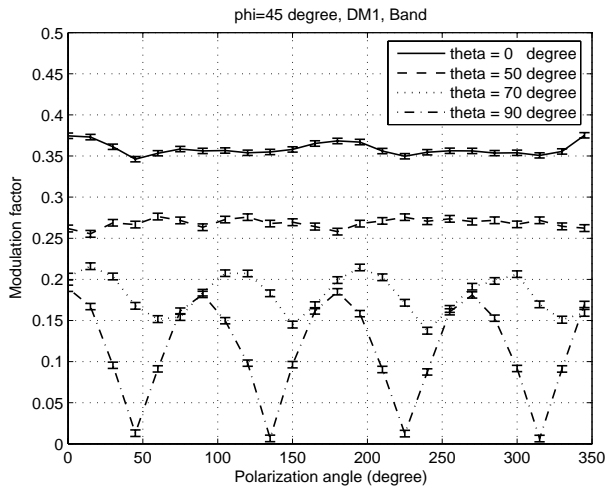

*Fig. 11 The modulation factor varies with the polarization angle of incident photons when the incidence angle phi=45 degree*

For simplification the average of the modulation factor over the polarization angle is used to calculate the MDP and estimate the performance. As shown in Fig. 12, the average of the modulation factor almost remains the same for different phi angle, while it decreases as the theta angle increases. The maximum of the modulation factor of POLAR can be as high as 0.35. Fig. 13 shows the effective area with respect to incident direction. The typical effective area reaches about 70 cm$^2$.

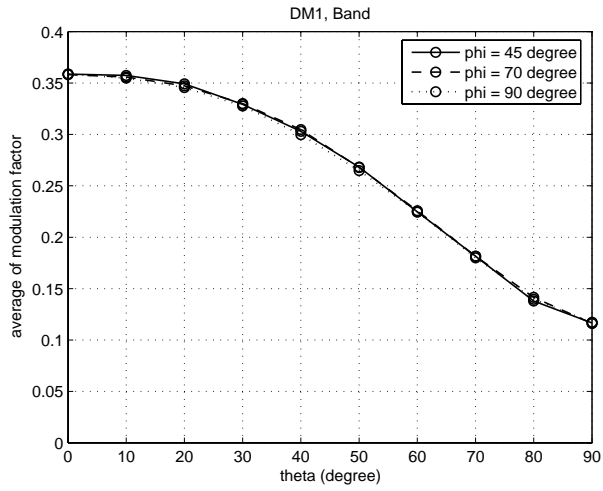

*Fig.12. The average of the modulation factor varies with the theta angle but remains almost constant with the phi angle*

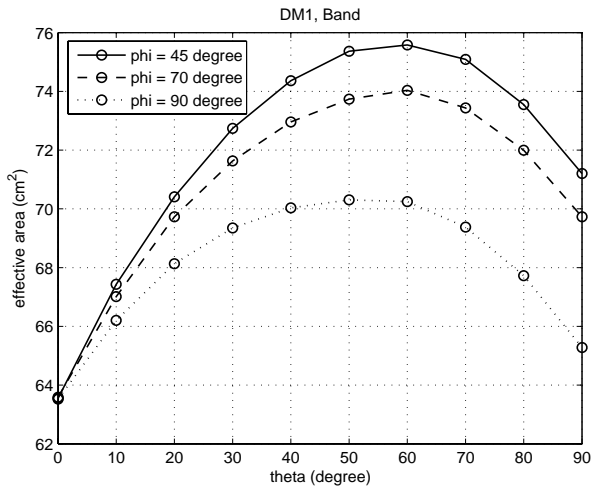

*Fig.13. The effective area varies with the direction of the incident photons.*

5.2 Background

There are several sources of background. The charged particle background can be eliminated almost completely by the trigger strategy with the upper energy cut. According to simulation, for electrons of 100 keV to 20 MeV, less than 0.1% will contribute to background, and for protons of 100 keV to 200 MeV, almost no background will be induced. The detailed flux of electrons and protons depends on the orbit of the spacecraft which is not available yet. Estimating for the most probable orbit of 350 km, the flux of electrons and protons outside of the South Atlantic Anomaly (SAA) is very low (less than several counts/cm$^2$/s), hence the contributing background is less than 10 counts/s. Consequently, in this paper the charged particle-induced background is not a concern.

Induced gamma ray flux depends very much on the orbit and the configuration of the spacecraft. Learning from the ISGRI instrument of Integral, this kind of background is very small compared to the cosmic x-ray background (CXB). In addition, the spacecraft structure of the Chinese Space Laboratory is not available. Therefore, simulations are emphasized on the background introduced by CXB. The spectra of CXB used here is showed below [40],

$$3-60 keV: 7.877E^{-0.29}e^{-E/41.13}\frac{keV}{keV\bullet cm^2\bullet s\bullet sr}$$

$$>60 keV: 0.0259(E/60)^{-5.5}+0.504(E/60)^{-1.58}+0.0288(E/60)^{-1.05}\frac{keV}{keV\bullet cm^2\bullet s\bullet sr} \quad (7)$$

In our simulation, the CXB comes from half of the sky isotropically since POLAR is supposed to point to the zenith in orbit and the contribution of the CXB from the backside is ignored because of the heavy Space Laboratory located under POLAR acting as a shielding structure. The azimuthal distribution is shown in Fig. 14. The modulation with a 90 degree periodicity is induced by the square geometry of POLAR. This result is very well consistent with the previous works [33]. As a rough prediction from simulation, the background rates will reach 300 counts/s and 470 counts/s for DM1 and DM2, respectively.

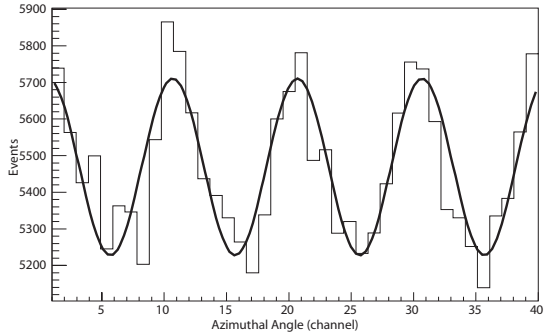

*Fig.14   The azimuthal distribution of background induced by CXB*

## 6 Performance

If the modulation factor and effective area are known for all possible incident directions as well as the background rate, and the direction, intensity and duration for the GRB are known then the MDP can be calculated according to the Equation 4. Generally, the simulated observation of POLAR can be fulfilled as follows:

a) Sample a GRB from the BATSE Current Gamma-Ray Burst Catalog [39] randomly and uniformly;

b) Choose the location of the GRB isotropically in the full sky;

c) Make sure that the GRB is located in the FOV of POLAR;

d) If the GRB lies in the FOV, get the modulation factor ($M_{100}$) and effective area according to the location of the GRB; All spectra of GRBs are assumed to follow the Band spectrum in Section 5.1 to simplify the simulation, which will introduce little difference into modulation factor and detection efficiency.

e) Read the intensity and duration of the GRB, calculate the MDP with the background rate.

f) Repeat a) to e), and count the number of GRB with a MDP not more than a certain value, for example, 5%, 10%, etc.

g) Taking the actual occurrence rate of GRBs into account, get the expected performance of POLAR in a one year observation (see Fig. 15).

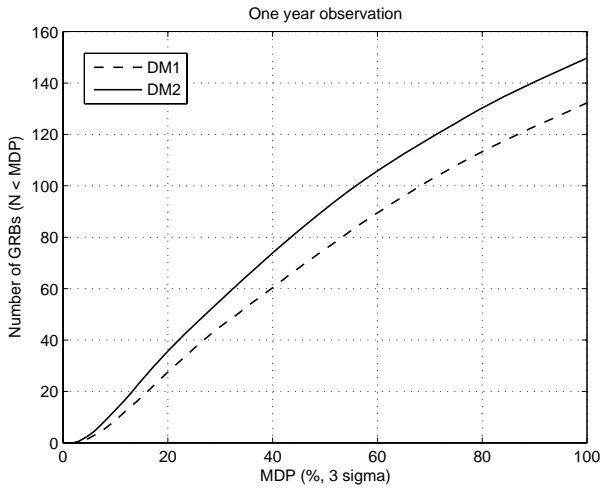

*Fig. 15 The expected performance of POLAR in a one year observation*

As showed in Fig. 15, for DM1, the expected number of GRBs with a MDP not more than 10% is 8.8 per year of observation, while this number is 12.8 for DM2. The polarization levels predicted by different models are summarized as follows: Synchrotron with ordered magnetic field model can produce linear polarization of 20%~70%, while synchrotron with small-scale random magnetic field model and Compton drag model will lead to < 20% typically. The maximum polarization can reach 100% for Compton drag model while 70% for synchrotron with small-scale random magnetic field model [41]. Consequently, the performance of POLAR is sufficient to distinguish these theoretical models for the prompt emission of GRBs.

## 7 Conclusions

As a very attractive payload for the Chinese Space Laboratory, the performance of POLAR, a dedicated polarimeter of GRB prompt emission, is evaluated by simulation in this paper, and about 9 GRBs/yr can be detected with MDP < 10% for DM1 and about 13 GRBs/yr for DM2, which demonstrates that the science goal of POLAR can be achieved after one year of observation. The simulation here is specially done for the case that POLAR will be aboard the Chinese Space Laboratory. The phenomenon of the modulation factor varying with the polarization angle of incident photons is revealed, which hasn't been investigated in any previous work. Moreover, the FOV is determined according to the actual placement of the Chinese Space Laboratory, which is very important to estimate how many GRBs POLAR can observe. However, the accuracy of the MDP calculated for the GRBs is limited by the assumption that all GRB spectra are fit by a Band function and by using a background rate that only included the background induced by the CXB. Other background sources, including charged particles, induced gamma rays of spacecraft and neutrons, will be taken into account once the configuration and orbit of the Chinese Space Laboratory are available.

**Acknowledgements**

The authors would like to thank the anonymous reviewer very much for many invaluable comments and suggestions to this paper. This work is supported by 973 Program 2009CB824800.